\documentstyle[aps,preprint]{revtex}
\draft

\author{Zafar Ahmed \\
Nuclear Physics Division, Bhabha Atomic Research Centre \\
Trombay, Bombay 400 085, India \\
zahmed@apsara.barc.ernet.in}
\title
{Pseudo-Hermiticity of Hamiltonians under imaginary shift
of the co-ordinate : real spectrum of complex potentials}
\date{\today}
\begin{document}
\maketitle
\begin{abstract}
We propose that the real spectrum and the orthogonality of the states for several
known complex potentials of both types, ${\cal PT}$-symmetric and
non-${\cal PT}$-symmetric can be understood in terms of currently proposed
$\eta$-pseudo-Hermiticity (Mostafazadeh, quant-ph/0107001) of a Hamiltonian,
provided the Hermitian linear automorphism, $\eta$, is introduced as $e^{-\theta p}$
which affects an {\it imaginary shift of the co-ordinate} : $e^{-\theta p}~ x
~e^{\theta p}  = x+i\theta$.
\end{abstract}
\vspace {.2 in}
Until year 1998 [1], Hermiticity of the Hamiltonian was supposed to be the
necessary condition for having real spectrum. A conjecture due to Bender and
Boettcher [1], has relaxed this condition in a very inspiring way by introducing
the concept of ${\cal PT}$-symmetric Hamiltonians. Here, ${\cal P}$ denotes
the parity operation (space reflection) : $x\rightarrow -x$ and ${\cal T}$
mimics the time-reversal : $i\rightarrow -i$. Let $\chi$ denote ${\cal PT}$
then if (i)- $\chi H \chi^{-1} = H$  and if (ii)- $\chi \Psi (x)=\pm 1 \Psi(x)$ the
eigenvalues will be real and complex conjugate pairs if the latter condition
is not satisfied. Several ${\cal PT}$-symmetric potentials have been witnessed
to hold real discrete spectrum [1-10]. A fully exactly solvable ${\cal PT}$-symmetric potential
model containing both the scenarios (real and complex conjugate pairs of
eigenvalues) is available [7]. Orthogonality conditions for the eigenstates
of such potentials have been proposed [7,11,12]. Recent conceptual developments
can be seen in Refs. [12-14].
\par However, a non-${\cal PT}$-symmetric complex Morse potential,
\begin{equation}
V(x)=(A+iB)^2\exp(-2x)-(2C+1)(A+iB)\exp(-x),
\end{equation}
which holds real discrete spectrum [10] remains a well known exception in this
regard. This complex potential (1) being exactly solvable, it becomes even more
interesting to find the underlying structure responsible for the real spectrum,
for a possible new direction.
The potentials discussed so far [1-10] can mainly be categorized as of two types
: $V_I(x)=\alpha V_e(x)+i\beta V_o(x)$ [1-7] ($e$: even, $o$: odd function) and
$V_{II}(x)=\alpha V_e(x-i\gamma)+i\beta V_o (x-i\gamma)$ [8-10].
Let us write a new category of potentials as $V_{III}(x)=\alpha V(x-\beta-
i\gamma)$, where $V(x)$ need not have a definite parity. Such potentials are
non-${\cal PT}$-symmetric, a few potentials of this category have been
generated [10] by group theoretic techniques and found to have real spectrum.
The complex Morse potential (1) is one such example. In these expressions the
parameters $\alpha$, $\beta$ and $\gamma$ are assumed to be real.
\par Currently, in a very interesting work [15], Mostafazadeh points out that
the potentials of the type (I,II) are actually ${\cal P}$-pseudo-Hermitian and
claims that the $\eta$-pseudo-Hermiticity
\begin{equation}
\eta~H~\eta^{-1}~=~H^\dagger,
\end{equation}
is the necessary condition for having real spectrum, where $\eta$ has been
referred to as {\it Hermitian linear automorphism}. Let ${\cal V}$ be an inner product
space, for two arbitrary elements $u$ and $v$, $\eta$ satisfies
\begin{equation}
<\eta u|v>~=~<u|\eta v>.
\end{equation}
In this Letter, we propose the imaginary shift of the co-ordinate
\begin{equation}
e^{-\theta p} x e^{\theta p}=x +i\theta,
\end{equation}
as a Hermitian linear automorphism, where $\theta$ is real and $p=-i{d\over dx}$,
$(\hbar = 1)$. \\
{\bf Proof :}
\begin{eqnarray}
x~e^{\theta p}-e^{\theta p} x &=& [x,e^{\theta p}] \nonumber\\
&=&\sum _{n=0}^{\infty} {\theta^n \over n!} [x,p^n] \nonumber\\
&=&\sum _{n=0}^{\infty} {in\theta^n p^{n-1} \over n!}\nonumber\\
&=&i\theta e^{\theta p}.
\end{eqnarray}
Multiplying from left on both sides by $e^{-\theta p}$, we prove the proposition
(4). The linear operator $e^{-\theta p}$ being Hermitian the condition (3) gets
automatically satisfied as $(e^{-\theta p}~\psi(x)) ^\dagger= \Psi^\ast (x)~
e^{-\theta p}$.
Here, it is worthwhile to recall that the usual linear operator,
$e^{-ipa}$, affecting a real shift in $x$ is unitary and not Hermitian.
The operator $e^{-\theta p}$ is endowed with the following important
properties :\\
\begin{mathletters}
A constant scalar $c$ (real or complex) commutes with $p$, so we have
\begin{equation}
e^{-\theta p}~c~e^{\theta p}~=~c.
\end{equation}
The linear momentum $p$ commutes with $e^{-\theta p}$ to remain unaffected
under imaginary shift of the co-ordinate. So we have
\begin{equation}
e^{-\theta p}~p~e^{\theta p}~=~p.
\end{equation}
By nothing that
$e^{-\theta p}~x^2~e^{\theta p}=e^{-\theta p}~x~e^{\theta p}
~e^{-\theta p}~x~e^{\theta p}~=~(x+i\theta)^2,$ the method of induction
leads to
\begin{equation}
e^{-\theta p}~x^n~e^{\theta p}~=~(x+i\theta)^n.
\end{equation}
A further generalization is by noting that the potential operator $V(x)$ can
be expanded in the powers of $x$, to write
\begin{equation}
e^{-\theta p}~V(x)~e^{\theta p}~=~V(x+i\theta).
\end{equation}
Reminding us of the Taylor series expansion, the operator $e^{-\theta p}$
would act on a wavefunction as
\begin{equation}
e^{-\theta p}~\Psi(x)~=~\Psi(x+i\theta).
\end{equation}
\end{mathletters}
The abovementioned ${\cal PT}$-symmetric potentials of the types $V_{I}(x)$ [1-7] and $V_{II}(x)$ [8-10]
have been discussed as being ${\cal P}$-pseudo-symmetric [15]. Nevertheless,
the complex Morse potential (1) and the potentials of the types $V_{III}(x)$
[10], despite entailing real spectrum require a Hermitian linear automorphism
for pseudo-Hermiticity to be invoked. Having introduced the properties of
$e^{-\theta p}$, we claim it to be the required Hermitian linear automorphism
for this exceptional class of the potentials. Notice that the
potential in Eq. (1) satisfies
\begin{equation}
e^{-\theta p}~V(x)~e^{\theta p} ~=~V(x+i\theta)~=~V^\ast(x),~\theta= [2\tan^{-1} (B/A)],
\end{equation}
rendering the Hamiltonian, $H=p^2+V(x)$, as pseudo-Hermitian (see Eq. [2])
under the imaginary shift of the co-ordinate. The real spectrum of the potential
(1) has been predicted by a group theoretic technique [10].
\par In the following, we propose to investigate the complex Morse potential (1) in the
light of pseudo-Hermiticity under imaginary shift of the co-ordinate as an illustrative
example. To this end, it is worthwhile to solve the Schr{\"o}dinger equation for
a general Morse potential
\begin{equation}
V(x)=V_1 \exp(-2x)- V_2 \exp(-x)
\end{equation}
using the first principles.
By assuming $\hbar=1=2m$, we write the Schr{\"o}dinger equation for (8)
\begin{equation}
{d^2\Psi(x) \over dx^2}+[E-V_1\exp(-2x) + V_2\exp(-x)]\Psi(x)=0.
\end{equation}
Next, a change of variable, namely $z=2\sqrt{V_1}\exp(-x)$ leads to
\begin{equation}
z^2{d^2\Psi(x) \over dz^2}+z{d\Psi(z) \over dz}+[E-{z^2 \over 4}
+{V_2 \over 2\sqrt{V_1}}z]\Psi(x)=0
\end{equation}
an important form which clearly indicates that the energy eigenvalues for a
general Morse potential will be a function of the effective parameter
${V_2 \over 2\sqrt{V_1}}$. Further, the fact that for the complex Morse
potential (1)
this effective parameter is real $(=C+1/2)$ explains the real spectrum to be
found in the sequel.
Next by using a transformation, $\Psi(z)=z^{\alpha}~ \exp(-z/2)~ F(z)$, in
Eq. (10) we get the confluent hypergeometric equation [16] as
\begin{equation}
zF^{\prime\prime}(z)+(2\alpha+1-z)~F^\prime(z)-(\alpha-C)~F(z)=0,~\alpha=\sqrt{-E}.
\end{equation}
We can write the solution of Eq. (10) in terms confluent hypergeometric function :
$\Psi(x)=z^\alpha~e^{-z \over 2}~_1F_1[\alpha-C, 2\alpha+1; z]$. The condition
that the series $_1F_1$ be truncated and become a polynomial gives the quantization
condition, $\alpha -C=-n, n=0,1,2..< C$, which in turn gives energy eigenvalues as
\begin{equation}
E_n=-(n-C)^2,~n=0,1,2,...< C = {V_2 \over 2\sqrt{V_1}}-1.
\end{equation}
It can also be checked that $\Psi_n(\pm \infty)=0$, as long as $n<C$. Finally,
the eigenfunctions can be re-written in terms of associated Laguerre polynomials as
\begin{equation}
\Psi_n(x)=z^{C-n}~e^{-z/2}~L^{2C-2n}_n(z),~z=2\sqrt{V_1} \exp(-x).
\end{equation}
The Schr{\"o}dinger equation (10) also being a Strum-Liouville equation
the orthogonality  condition, $\int_{-\infty}^{\infty} \Psi_m(x)~ \Psi_n(x) dx~
=~ N^2_n \delta_{m,n}$, ought to be satisfied. However, surprisingly, in the
mathematical handbooks [16,17], we do not find the interesting ensuing integral,
namely
\begin{equation}
\int_{0}^{\infty} z^{2c-(m+n+1)}~e^{-z}~L^{2c-2m}_m(z)~L^{2c-2n}_n(z)~dz~=~N^2_n~\delta_{m,n}.
\end{equation}
We have, therefore, checked the result (14) using $NIntegrate$ of $Mathematica$ for
several cases. We think it remains desirable to prove the result in Eq. (14) analytically.
\par Having confirmed the real spectrum of the pseudo-Hermitian potential (1) under
the imaginary shift of the co-ordinate (4), we now show that it conforms to the
proposed definition of $\eta$-orthogonality [15], namely
\begin{equation}
(E^\ast_1-E_2)\int_{-\infty}^{\infty} \Psi^\ast_1 (x)~ \eta~ \Psi_2(x)~dx =0.
\end{equation}
The ${\cal PT}$-orthogonality has earlier [7,12] been proposed as
\begin{equation}
(E^\ast_1-E_2)\int_{-\infty}^{\infty} \Psi^{PT}_1 (x)~ \Psi_2(x)~dx = 0,
\end{equation}
which is not different from (15) in case the Hamiltonian is
${\cal P}$-pseudo-Hermitian.
Let us notice the operation of imaginary shift of the co-ordinate on
$z(x)(=2(A+iB)~e^{-x})$ using Eq. (6d) for the special case of complex Morse
potential (1) as
\begin{equation}
e^{-\theta p}~z(x)~e^{\theta p}=z(x+i\theta)=2(A+iB)~e^{-(x+i\theta)}=2(A+iB)
{A-iB \over A+iB}~e^{-x}~=~z^\ast(x).
\end{equation}
Similarly, using the property in Eq. (6e) for an arbitrary complex eigenstate
from Eq. (10), we find that
$e^{-\theta p} \Psi_n(z) = \Psi_n(z^\ast) = \Psi^\ast_n(x)$ to display the
orthogonality effectively as $\int_{-\infty}^{\infty} \Psi^\ast_m(x) \Psi^\ast_n(x)dx
= N^2_n \delta_{m,n}$, in conformity with the condition (15) and the
mathematical result (14).
\par Interestingly, the potentials, $V_I(x)$ [1-7] and $V_{II}(x)$ [8-10] are
both ${\cal PT}$-symmetric and ${\cal P}$-Pseudo-Hermitian as well. However,
we assert that
the $III$ category of potentials, $V_{III}(x)=\alpha V(x-\beta-i\gamma)$ which
are non-${\cal PT}$-symmetric are only pseudo-Hermitian under the imaginary shift
of the co-ordinate ($x \rightarrow x+2i\gamma$) :
\begin{equation}
e^{-2\gamma p}~\alpha V(x-\beta-i\gamma)~e^{2\gamma p}~=~\alpha V(x+2i\gamma-\beta-i\gamma)
~=~\alpha  V^\ast(x-\beta-i\gamma).
\end{equation}
The wavefunctions will be such that $e^{-2\gamma p} \Psi_n(x-\beta-i\gamma)~
=~\Psi(x-\beta+i\gamma)$ to satisfy the orthogonality condition (15),
given the fact that the states, $\Psi_1(x)$ and $\Psi_2(x)$, of the real
potential, $V(x)$, are orthogonal in $[-\infty,\infty]$.
We point out that in Ref. [6], the potential, $V(x)=[z\cosh(2x)-iM]^2$
is actually pseudo-Hermitian under $x\rightarrow x+i\pi/2$.
If, $\hbar=1=m$, three Harmonic-oscillator
potentials : $V^{HO}_1(x)={1 \over 2} x^2, V^{HO}_2(x)= {1 \over 2}(x-i\gamma)^2$
and $V^{HO}_3(x)={1 \over 2}(x-\beta-i\gamma)^2$ will have the same real
spectrum, $E_n=n+1/2$, with different sets of eigenfunctions.
The potential $V^{HO}_1$ is Hermitian, $V^{HO}_2$ is both ${\cal PT}$-symmetric
and pseudo-Hermitian as well, and $V^{HO}_3$ is only pseudo-Hermitian. Both the
times the pseudo-Hermiticity is under the imaginary shift of the co-ordinate (4).
\par When $\hbar=1=2m$, three Eckart potentials $V^E_1(x)=-\alpha~
{\rm sech^2}(x), V^E_2(x)= -\alpha~ {\rm sech^2}(x-i\gamma)$ and $V^E_3(x)
=-\alpha~ {\rm sech^2}(x- \beta-i\gamma)$, will all have a common real
spectrum, $E_n=-[n+1/2- \sqrt{\alpha+1/4}]^2,~n=0,
1,2...<\sqrt{\alpha+1/4}$ with different sets of eigenstates. Once again,
$V^E_1$ is Hermitian, $V^E_2$ is both ${\cal PT}$-symmetric and pseudo-
Hermitian and $V^E_3$ is only pseudo-Hermitian. Once again, both the times the pseudo-Hermiticity
is under imaginary shift of the co-ordinate (4). Apparently similar looking potentials $V^E_2$ and $V^E_3$
can be made to look entirely different from $V^E_1(x)$, in order to visualize
and analyze them like their partner $V^E_1(x)$, if their real and imaginary
parts are separated out.
\par Further, we can cite
the exactly solvable well known potentials P{\"o}schal-Teller-II and
generalized P{\"o}schal-Teller [9,10] which can contribute the
pseudo-Hermitian potentials in the proposed sense (4). They are $V^{PT-II}
(x-\beta-i\gamma)$ and $V^{GPT}(x-\beta-i\gamma)$; they will have a common
real spectrum as that of $V^{PT-II}(x)$ and $V^{GPT}(x)$, respectively.
\par The formalism [15] does not claim the $\eta-$pseudo-Hermiticity to be
{\it sufficient}, however, it has been claimed to be the {\it necessary}
condition on the complex potential for having the real spectrum. We remark
that the latter issue is not without a practical difficulty. The main problem
in this regard is to actually identify the Hermitian linear automorphism,
$\eta$. In some instances it might be apparent, in others it might even
be elusive for the pseudo-Hermiticity to be talked about. We feel that the
identification of the imaginary shift of the co-ordinate, $e^{-\theta p}$
to act as $\eta$ is the main contribution of the present Letter. In this light, several
complex potentials of both the types, ${\cal PT}$-symmetric and
non-${\cal PT}$-symmetric, entailing real spectrum have been argued to be
pseudo-Hermitian. The complex Morse potential turns out to be a bit more
interesting instance of pseudo-Hermiticity under imaginary shift of the
co-ordinate. Further, introduction or identification of a new Hermitian linear
automorphism, $\eta$, will help in generating a new class of non-Hermitian Hamiltonians
having real spectrum.
\section*{References :}
\begin{enumerate}
\item C.M. Bender and S. Boettcher, Phys. Rev. Lett.  80 (1998) 5243.
\item C.M. Bender, S. Boettcher, P.N. Meisinger, J.Math. Phys. 40 (1999) 2201.
\item F. Cannata, G. Junker, J. Trost, Phys. Lett. A 246 (1998) 219.
\item M. Znojil; J. Phys. A: Math. Gen. 33 (2000) L61.
\item B. Bagchi and R. Roychoudhury, J. Phys. A: Gen. Math. 33 (2000) L1.
\item A. Khare and B.P. Mandal, Phys. Lett. A 272 (2000) 53.
\item Z. Ahmed, Phys. Lett.  282  (2001) 343 and an Addendum to appear.
\item M. Znojil, Phys. Lett. A  259 (1999) 220.
\item G. Levai and M. Znojil, J. Phys. A : Math. Gen.  33 (2000) 7165.
\item B. Bagchi and C. Quesne, Phys. Lett. A  273 (2000) 285.
\item C.M. Bender, F. Cooper, P.N. Meisinger, V.M. Savage, Phys. Lett. A 259
(1999) 224.
\item G.S. Japaridze `Space of state vectors in ${\cal PT}$-symmetrical
quantum mechanics', LANL Archives : quant-ph/0104077.
\item M. Znojil,`Conservation of pseudo norm in ${\cal PT}$-symmetric
quantum mechanics' LANL Archives : math-ph/0104012.
\item R. Kretschmer,and L. Szymanowski `The interpretation of quantum
mechanical models with non-Hermitian Hamiltonians and real spectra', LANL
Archives : quant-ph/0105054.
\item A. Mostafazadeh, `Pseudo-Hermiticity versus ${\cal PT}$-symmetry :
The structure responsible for the reality of the spectrum of a non-Hermitian
Hamiltonian', LANL Archives : quant-ph/0107001.
\item I.S. Garadshteyn and I.M. Ryzhik, {\it Tables of Integrals Series and
Products}, $4^{th}$ edn., translation edited by A. Jeffrey (Accademic,
New York, 1965).
\item A. Erdelyi, W. Magnus, F. Oberhettinger, F. G. Tricomi, {\it Tables of
Integral Transforms} vol. 2 (McGraw-HILL, New York, 1954) pp. 292-294.
\end{enumerate}
\end{document}